\documentclass{article}

\usepackage{arxiv}

\usepackage[utf8]{inputenc} 
\usepackage[T1]{fontenc}    
\usepackage{hyperref}       
\usepackage{url}            
\usepackage{booktabs}       
\usepackage{amsfonts}       
\usepackage{nicefrac}       
\usepackage{microtype}      
\usepackage{lipsum}
\usepackage{graphicx}
\usepackage{subfigure}
\usepackage{amssymb}
\usepackage{amsmath}
\usepackage{mathtools}
\providecommand{\abs}[1]{\lvert#1\rvert}

\graphicspath{ {./images/} }

\title{Human-Centered Interactive Anonymization for Privacy-Preserving Machine Learning: A Case for Human-Guided k-Anonymity}

\author{
 Sri Harsha Gajavalli \\
  School of Coumputing and Augmented Intelligence\\
  Arizona State University\\
  Tempe, AZ 85281 \\
  \texttt{gsh@asu.edu}
}

\begin{document}
\maketitle

\author{\IEEEauthorblockN{Sri Gajavalli}
\IEEEauthorblockA{\textit{sriharsha.g15@iiits.in}}
}	
\maketitle

\begin{abstract}
Privacy-preserving machine learning (ML) seeks to balance data utility and privacy, especially as regulations like the GDPR mandate the anonymization of personal data for ML applications. Conventional anonymization approaches often reduce data utility due to indiscriminate generalization or suppression of data attributes. In this study, we propose an interactive approach that incorporates human input into the k-anonymization process, enabling domain experts to guide attribute preservation based on contextual importance. Using the UCI Adult dataset, we compare classification outcomes of interactive human-influenced anonymization with traditional, fully automated methods. Our results show that human input can enhance data utility in some cases, although results vary across tasks and settings. We discuss limitations of our approach and suggest potential areas for improved interactive frameworks in privacy-aware ML.

\end{abstract}

\renewcommand{\thesubfigure}{\thefigure.\arabic{subfigure}}
\makeatletter
\renewcommand{\p@subfigure}{}
\renewcommand{\@thesubfigure}{\thesubfigure:\hskip\subfiglabelskip}
\makeatother

\section{Introduction and Rationale}
\label{sect:intro_moti}

Across various domains in today’s data-centric world, technological advancements rely heavily on extracting insights from data, including mining information from diverse sources and analyzing personal data. The latter plays a critical role in fostering business intelligence and delivering tailored services, which are in high demand in contemporary society. However, achieving these objectives often necessitates the sharing, linking, and systematic processing of personal data from heterogeneous origins, which introduces significant risks of exposure. These risks can range from minor inconveniences, such as revealing a social media profile, to severe consequences, like the unintended disclosure of sensitive health data to unauthorized parties.

To address such challenges, governments worldwide are considering or have already implemented legislation to regulate personal data management. A notable example is the European General Data Protection Regulation (\textit{GDPR}), effective from June 1, 2018, which grants individuals the \textit{right to be forgotten}. This provision mandates organizations to delete an individual’s personal data upon request, unless the data has been anonymized prior to analysis. This directive has brought attention to Privacy-Aware Machine Learning (PaML), which involves applying machine learning techniques exclusively to anonymized datasets. Anonymization can be achieved through methods such as perturbing data by introducing noise (e.g., \textit{differential privacy} \cite{dwork2008differential}) or grouping data into equivalence classes (\textit{k-anonymity} \cite{Sweeney:2002:k-Anonymity}), the latter of which has become widely adopted in the industry.

The initial framework of \textit{k-anonymity} has since been refined with enhancements like \textit{l-diversity} \cite{MachanavajjhalaEtAl:2007:l-Diversity}, which ensures that each group contains a minimum diversity of sensitive values; \textit{t-closeness} \cite{LiEtAl:2007:t-closeness}, which limits the divergence between local and global distributions of sensitive values; and \textit{delta-presence} \cite{NergizClifton:2010:Delta-Presence}, which accounts for an adversary’s background knowledge. While these advancements offer intriguing perspectives, for this study—focused on contrasting interactive machine learning algorithms with automated methods—we restrict our exploration to \textit{k-anonymity}.

Building upon earlier work \cite{malle2016right, MalleKieseHolzinger:2017:DoNotDisturb}, which examined binary classification accuracy on partially modified versus fully anonymized datasets, this study proposes the use of interactive machine learning as a novel approach to ($k$-)anonymization.

\section{k-Anonymity}
\label{sect:k_anon}

Anonymization of datasets is a cornerstone of modern data privacy, particularly as datasets often include attributes that could potentially reveal sensitive or identifiable information. These attributes typically fall into three distinct categories:  

\begin{itemize}
    \item \textbf{Direct identifiers} are data fields that unequivocally identify an individual without requiring any additional context or cross-referencing. Examples include unique identifiers such as email addresses, phone numbers, or social security numbers (SSNs). These attributes pose a direct threat to individual privacy and are therefore the first to be excluded or masked during the anonymization process. Their removal ensures that no individual can be explicitly pinpointed from the dataset.

    \item \textbf{Sensitive attributes}, often termed the "payload" of a dataset, consist of critical information essential for the dataset’s intended analytical or research goals. Examples include medical conditions, salary ranges, or political affiliations. These attributes are the focal point of analysis and cannot be eliminated or heavily generalized without compromising the utility of the dataset. For instance, in a medical study, preserving the specific disease classifications is vital for accurate research outcomes. Therefore, sensitive attributes require careful handling to ensure both privacy and analytical integrity.

    \item \textbf{Quasi-identifiers (QIs)} are attributes that do not directly identify individuals but, when combined with external information, could lead to re-identification. For example, \cite{sweeney2002k} demonstrated that 87\% of U.S. citizens could be uniquely identified using just three quasi-identifiers: \textit{zip code}, \textit{gender}, and \textit{date of birth}. These attributes are particularly challenging because they often carry valuable insights. For instance, zip codes provide critical information for public health studies, such as identifying disease clusters or targeting vaccination campaigns. Consequently, QIs are typically subjected to generalization rather than outright removal, allowing researchers to maintain data utility while mitigating re-identification risks.
\end{itemize}

\textit{k-Anonymity}, introduced by \cite{Samarati:2001:kAnonymity}, provides a structured solution to privacy concerns by ensuring that every record in a dataset is indistinguishable from at least $k-1$ other records based on quasi-identifiers. This is achieved by organizing records into equivalence classes, where all records within a class share identical generalized quasi-identifier values.  

Generalization is a core mechanism of $k$-anonymization. It transforms specific data values into broader categories, thereby reducing the risk of identification while preserving dataset usability. For instance, a dataset containing the ZIP codes '8010,' '8045,' and '8500' could undergo multiple stages of generalization: initially grouped as '80,' then further abstracted to '8,' and eventually generalized. Each level of abstraction reflects a trade-off between privacy and utility, with higher levels of generalization providing greater anonymity but potentially diminishing analytical precision.  

Beyond generalization, other techniques such as suppression (removing specific data points) and noise addition (introducing small random errors) are sometimes employed to achieve $k$-anonymity. However, these methods must be carefully balanced to avoid excessive information loss, which can render the anonymized dataset unsuitable for its intended purpose.  

\section{Interactive Machine Learning}
\label{sect:iML}

Interactive Machine Learning (iML) represents a transformative approach to data processing, where algorithms iteratively improve their performance through feedback from an external \textit{oracle}, such as a human expert or user. Unlike traditional machine learning systems, which rely solely on predefined training datasets, iML systems adapt dynamically by incorporating insights provided during the learning process. This makes them particularly effective for applications requiring domain expertise, such as personalized medicine, financial forecasting, and data anonymization \cite{Holzinger:2016:iML}.  

A key advantage of iML is its ability to address computationally intensive problems, where exhaustive algorithmic approaches become impractical. Many real-world problems exhibit exponential or super-exponential complexity, meaning that brute-force solutions are infeasible. In such cases, human intuition and domain knowledge can significantly enhance the efficiency and accuracy of machine learning models \cite{iMLExperiment}. For example, a human expert can quickly identify meaningful patterns or rule out irrelevant data points, accelerating the learning process.  

In the context of data anonymization, existing systems have made strides toward incorporating user input. For instance, \cite{Moque2012} introduced a framework that allows users to configure the $k$-anonymity parameter and analyze the resulting information loss. However, this system operates in a batch mode, requiring users to wait for the entire anonymization process to complete before providing feedback. Similarly, the Cornell Anonymization Toolkit (CAT) \cite{Xiao2009} collects user feedback post hoc, limiting its ability to adapt dynamically during the learning phase.  

Our approach, by contrast, emphasizes real-time interaction. During the anonymization process, users can adjust algorithmic parameters and immediately observe the impact on metrics such as information loss and equivalence class size. This iterative feedback loop enables a more adaptive and responsive system, ensuring that user preferences are continuously incorporated into the anonymization process.  

Another notable contribution in the field comes from \cite{Loh2010}, which proposed user-defined constraints for attribute generalization and utilized domain-specific ontologies to construct hierarchical structures. While this represents a step toward semi-interactive anonymization, it still lacks the dynamic, real-time feedback loop that characterizes true iML systems.  

Beyond data privacy, iML has found applications across diverse domains. For example, in the medical field, iML has been used to study protein interactions, identify clusters in genomic data \cite{Amershi2014}, and assist in precision medicine by tailoring treatments to individual patients. In social network analysis, it has enabled on-demand group formation based on user-defined criteria \cite{Amershi2012}. Even in creative fields, iML has facilitated innovative projects, such as mapping human gestures to real-time music generation \cite{Fiebrink2009}. These examples underscore the versatility of iML and its potential to bridge the gap between computational power and human expertise.  

In conclusion, iML offers a promising framework for addressing the challenges of modern data anonymization. By combining the adaptability of machine learning with the nuanced judgment of human experts, iML systems can achieve a balance between privacy protection and data utility that would be difficult to achieve with traditional methods alone.
.

\section{Experiments}
\label{sect:experiments}

This section provides a detailed account of our experimental setup, covering the overarching iML framework, the dataset selection process, the anonymization algorithm employed, and the systematic pipeline used to produce and evaluate results. Our goal was to assess how interactive weight adjustments, guided by human oracles, influence anonymization outcomes in terms of utility and privacy preservation.

\subsection{General Setting}
\label{sect:setting}

The primary aim of our experiments was to investigate the influence of varying attribute importance weights—representing user preferences for specific quasi-identifiers—on the anonymization process. For example, in a dataset used for studying regional disease outbreaks, attributes like \textit{ZIP code} may be prioritized over less critical features such as \textit{occupation} or \textit{gender}. Alternatively, in a dataset tailored for dermatological research, \textit{race} might hold significant importance, while \textit{ZIP code} could be of secondary relevance.

\begin{figure*}[h]
	\begin{center}
		\vspace{-1.0cm}
		\hspace*{-0.8cm}
		\includegraphics[width=0.8\textwidth]{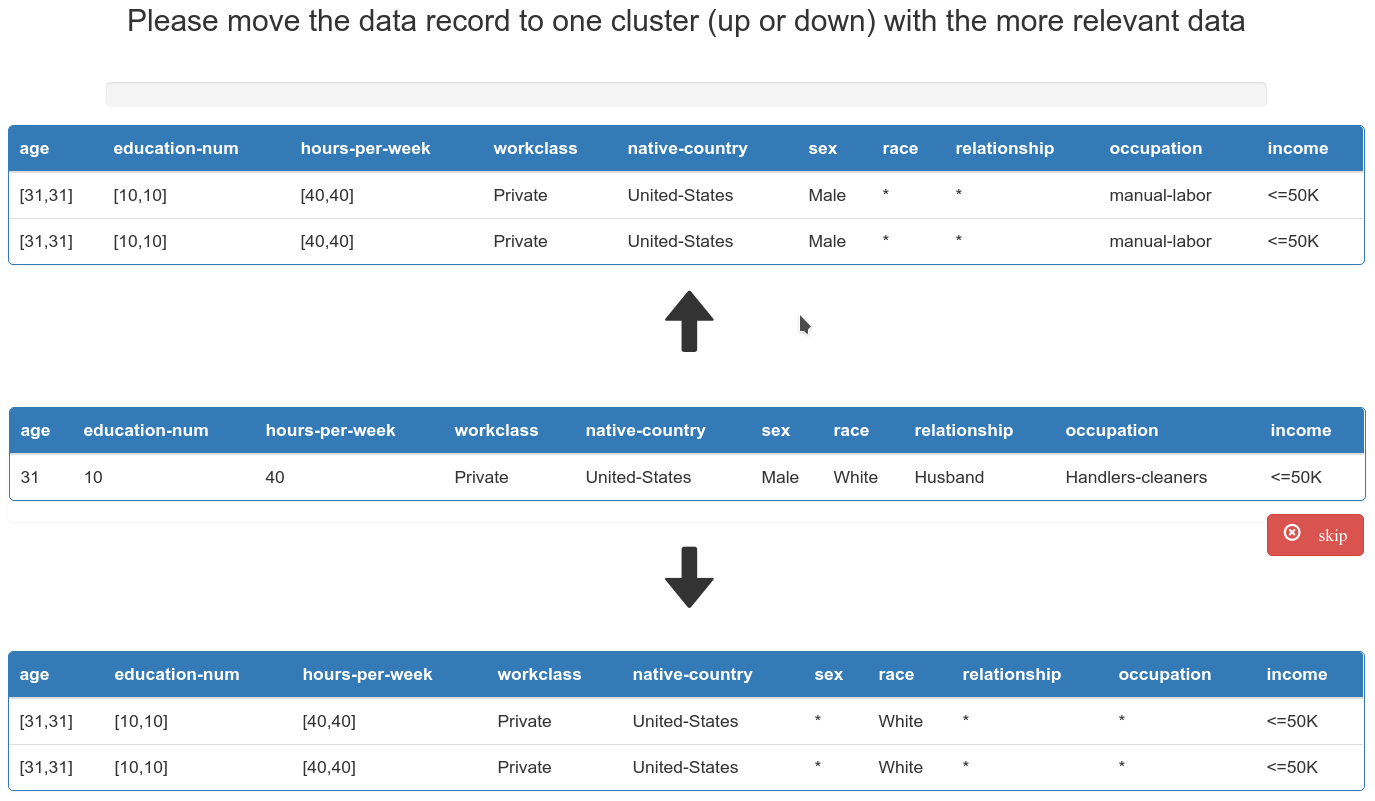}
		
		\vspace{0cm}
		\hspace*{-.8cm}
		\includegraphics[width=0.8\textwidth]{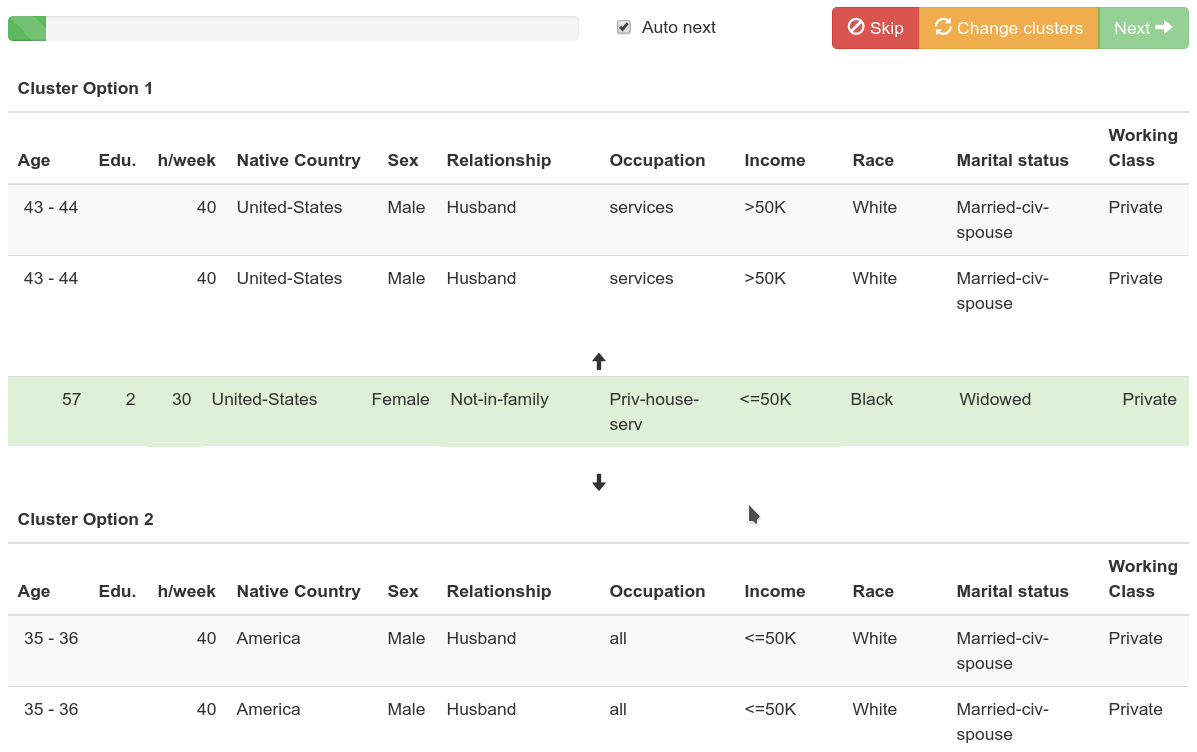}
		\caption{Illustration of two iterations of the iML interface for interactive anonymization.}
		\label{fig:iml_UI}
	\end{center}
\end{figure*}

Our study categorized weight configurations into three distinct approaches:
\begin{enumerate}
    \item \textbf{Equal Weighting:} All quasi-identifiers were treated with uniform importance, creating a baseline scenario devoid of user input.
    \item \textbf{User-Defined Bias:} A sliding-scale mechanism allowed participants to assign varying weights to attributes based on perceived relevance.
    \item \textbf{Interactive Machine Learning (iML):} Users actively participated in clustering data into partially anonymized subsets, providing direct feedback during the anonymization process. This approach facilitated real-time adjustments to the algorithm’s parameters and enabled a personalized balance between data utility and privacy (see Figure~\ref{fig:iml_UI}).
\end{enumerate}

These configurations were tested across three classification targets—\textit{income}, \textit{education level}, and \textit{marital status}—to evaluate how anonymization strategies affected predictive performance.

\subsection{Dataset Description}
\label{ssect:data}

We utilized the widely recognized Adults dataset from the UCI Machine Learning Repository, originally derived from 1994 U.S. Census Bureau data. This dataset comprises over 48,000 individual records, making it a robust choice for anonymization research. After preprocessing, we focused on two subsets:
\begin{itemize}
    \item A reduced dataset of 500 fully complete rows was used for testing the iML methodology, ensuring manageable interaction during the iterative learning process.
    \item A larger subset of 3,000 records was employed to generate 775 anonymized datasets for classifier evaluations. These datasets were produced using varying $k$-anonymity thresholds to analyze the trade-offs between privacy and utility.
\end{itemize}

To optimize the dataset for our study, we excluded the following attributes:
\begin{itemize}
    \item \textit{capital-gain} and \textit{capital-loss}, due to their highly skewed distributions, which reduced their interpretability for user interactions.
    \item \textit{fnlwgt}, a weighting factor irrelevant to the anonymization process.
    \item \textit{education}, which was redundant as it was already represented by the numeric attribute \textit{education\_num}.
\end{itemize}

The final dataset consisted of key attributes like \textit{age}, \textit{race}, \textit{gender}, \textit{ZIP code}, and \textit{income}, each carrying varying levels of significance depending on the experimental objective.

\subsection{Anonymization Algorithm}
\label{ssect:algorithm}

For implementing anonymization, we adopted \textit{SaNGreeA} (Social Network Greedy Clustering), a robust algorithm introduced by \cite{campan2009data}. This algorithm’s flexibility and computational efficiency made it particularly suitable for integration with our iML framework. 

\textit{SaNGreeA} operates as a greedy clustering algorithm with a complexity of $O(n^2)$, iteratively grouping records into clusters that satisfy the $k$-anonymity criterion. It minimizes General Information Loss (GIL), a metric quantifying the degree of abstraction introduced during generalization. GIL accounts for both continuous and categorical attributes and is defined as follows:

\begin{align*}
\text{GIL}(cl) &= \abs{cl} \cdot \Bigg( 
\sum_{j=1}^{s} 
\frac{\text{range(gen}(cl)[N_j])}{\text{range}(X[N_j])} \\
&\quad + \sum_{j=1}^{t} 
\frac{\text{height}(\Lambda(\text{gen}(cl)[C_j]))}{\text{height}(H_{C_j})} 
\Bigg)
\end{align*}

Where:  
- $\abs{cl}$ denotes the size of a cluster.  
- $\text{range}([i_1, i_2])$ is the size of the interval for numeric attributes.  
- $\Lambda(w), w \in H_{C_j}$ is the sub-hierarchy for categorical attributes rooted at $w$.  
- $\text{height}(H_{C_j})$ represents the height of the categorical hierarchy.  

\subsection{Processing Pipeline and Results Evaluation}
\label{ssect:process}

To evaluate the impact of iML, we developed a systematic pipeline comprising the following steps:
\begin{enumerate}
    \item \textbf{Dataset Preparation:} We selected the first 5,000 rows of the processed dataset and applied $k$-anonymization for $k$ values ranging from 5 to 200, generating 774 anonymized versions of the dataset.
    \item \textbf{Classifier Testing:} Anonymized datasets were evaluated using four machine learning classifiers—\textit{linear SVC}, \textit{logistic regression}, \textit{gradient boosting}, and \textit{random forest}. Each classifier was tasked with predicting three target attributes: \textit{income}, \textit{education level}, and \textit{marital status}.
    \item \textbf{Performance Metrics:} For each combination of target attribute and weight configuration (\textit{equal}, \textit{bias}, \textit{iML}), we averaged the classifier results. Metrics included accuracy, precision, recall, and F1 scores to assess the utility of anonymized datasets.
    \item \textbf{Visualization and Analysis:} Results were visualized as plots, where each point represented an anonymized dataset. The leftmost points corresponded to the original (non-anonymized) dataset, while the rest illustrated the trade-offs at increasing $k$ levels.
\end{enumerate}

Our results demonstrated that the iML approach consistently outperformed both baseline methods (\textit{equal} and \textit{bias weighting}) in achieving a favorable balance between privacy and utility. Specifically, the interactive feedback mechanism allowed participants to retain the most relevant attributes for the classification task while minimizing information loss. These findings highlight the potential of iML as a dynamic and user-centric approach to data anonymization.

\section{Results \& Discussion}
\label{sect:results}

Based on the findings from our prior investigation into PaML \cite{malle2016right} \cite{MalleKieseHolzinger:2017:DoNotDisturb}, we anticipated classifier performance to align with a $1/x$ curve as the values of $k$ increased. However, this trend was only partially observed. Specifically, for the targets \textit{education} and \textit{income}, no definitive trend emerged across weight categories, with performance varying depending on the particular factor of $k$.

\begin{figure}[h]
\begin{center}
\includegraphics[width=0.8\linewidth]{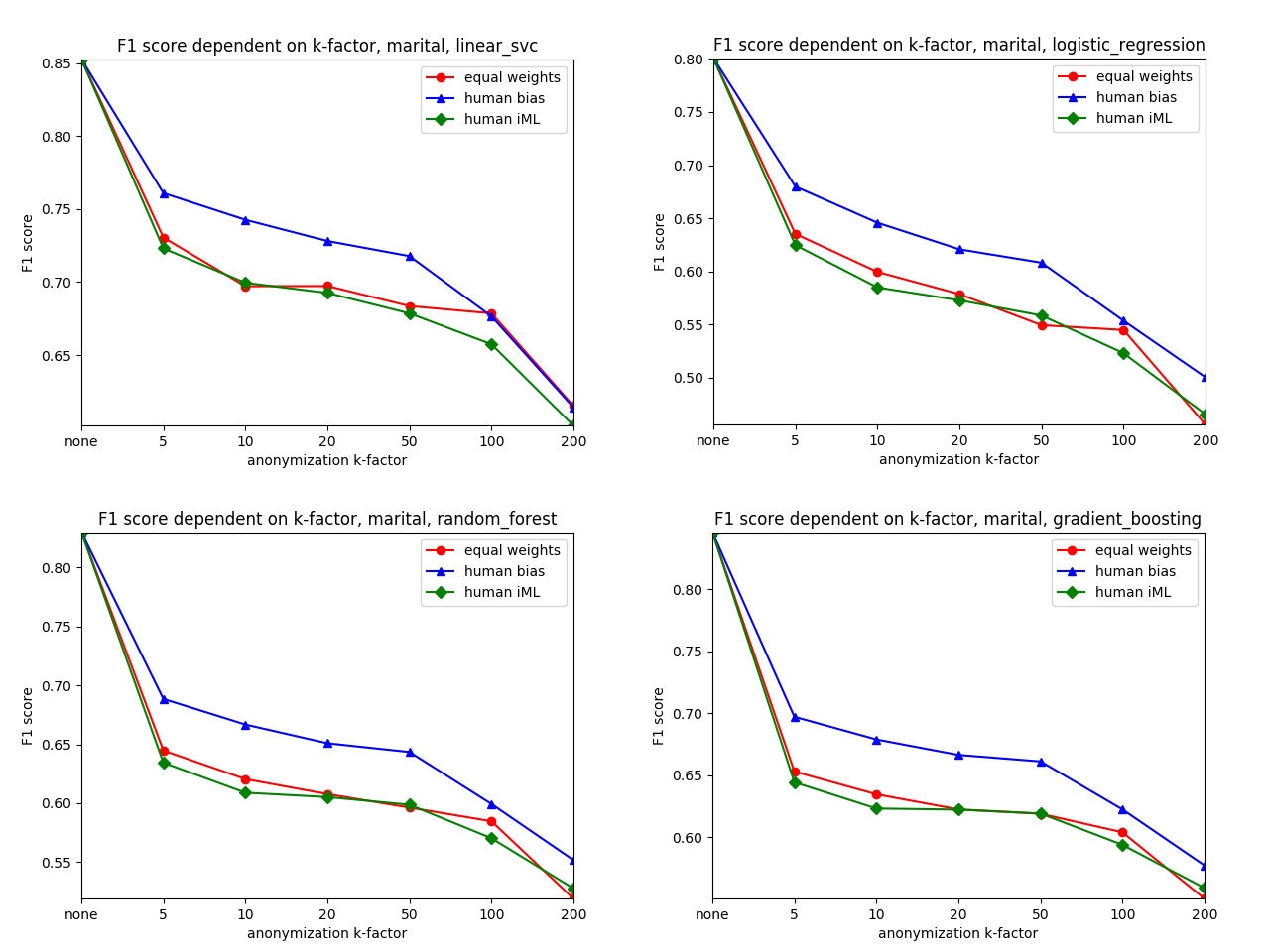}
\caption{Outcomes for the target \textit{marital status}: human bias consistently outperforms uniform weighting and human-algorithm interaction.}
\label{fig:results_marital}
\end{center}
\end{figure}

The target \textit{marital status} demonstrated the most consistent outcomes, with human-derived weights surpassing both uniform weights and user-guided interactions (Figure~\ref{fig:results_marital}). This may be attributed to a pronounced link between the attributes \textit{marital-status} and \textit{relationship} in the dataset, leading participants to consciously overemphasize the latter attribute. The relatively weaker performance of iML in this scenario, observed throughout our results, is discussed later in greater detail.

\begin{figure}[h]
\begin{center}
\includegraphics[width=0.8\linewidth]{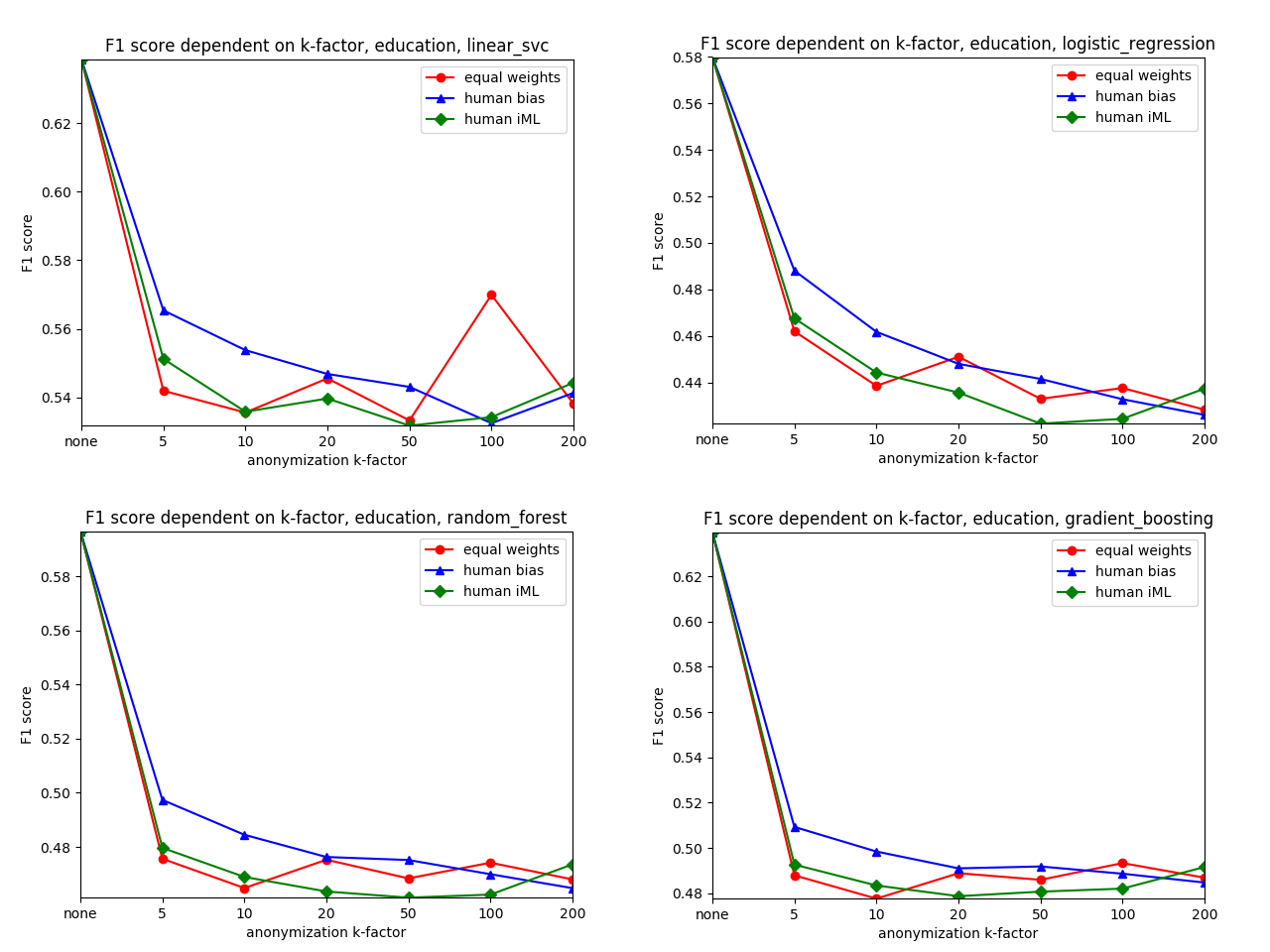}
\caption{Outcomes for the target \textit{education}: human bias generally outperforms other methods, albeit less consistently than for \textit{marital status}.}
\label{fig:results_education}
\end{center}
\end{figure}

For the target \textit{education}, human-derived weights generally outperformed iML-calculated weights, with uniform weights prevailing at higher $k$ values (Figure~\ref{fig:results_education}). One possible explanation is that misleading cues, such as \textit{income} or \textit{working hours}, might have influenced predictions. However, this does not fully account for the discrepancy between human bias and iML-based outcomes. Additionally, the performance on this target was substantially lower compared to the other scenarios, potentially reducing the significance of the observed gaps.

\begin{figure}[h]
\begin{center}
\includegraphics[width=0.8\linewidth]{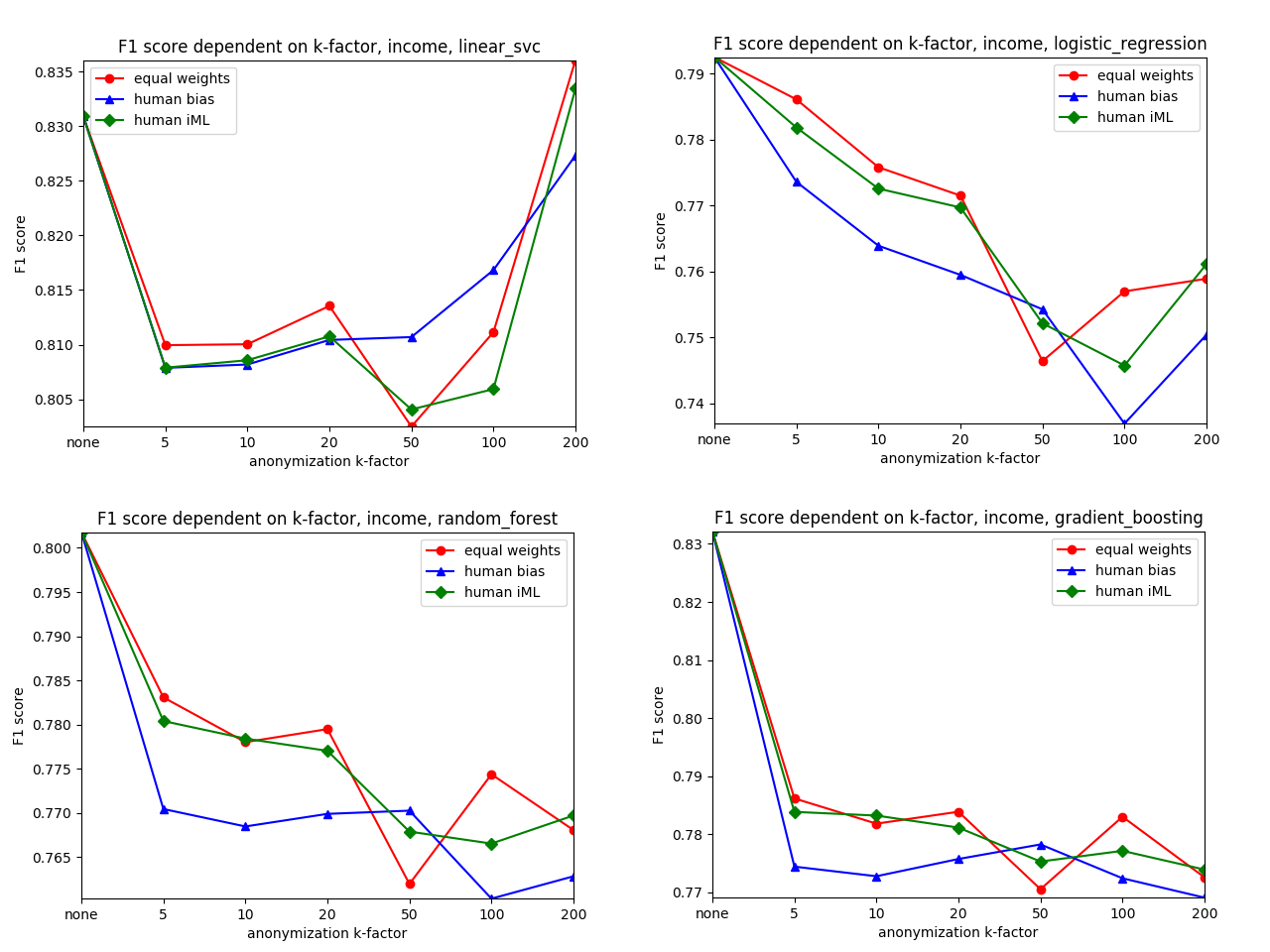}
\caption{Outcomes for the target \textit{income}: iML-based weights outperform human bias in some cases but fail to surpass the uniform weighting approach.}
\label{fig:results_income}
\end{center}
\end{figure}

For the \textit{income} target, the hierarchy between human bias and iML was occasionally reversed, but both typically underperformed compared to uniform weights (Figure~\ref{fig:results_income}). This is notable given that \textit{income} was the only binary classification task in our study, which might have given humans a slight advantage. However, susceptibility to stereotypes regarding factors like gender, race, or marital status could explain the decline in human bias performance.

The inability of iML to significantly outperform both human bias and uniform weighting can likely be attributed to the constraints of our experimental design. To enable real-time interaction, we opted for a simplified anonymization algorithm with $O(n^2)$ runtime, restricting the dataset to a mere 500 rows (1\% of the original dataset). This likely accelerated generalization and led to premature suppression, reducing the scope for effective clustering. Additionally, some users may not have engaged seriously with the experiment, a limitation that could be addressed by selecting more committed participants or employing a more engaging application domain.

Finally, it is intriguing that a $k$ factor as high as 200 still produced comparable or occasionally improved outcomes, despite its apparent absurdity.

\section{Open Problems \& Future Challenges}
\label{sect:op_fc}

The field of iML for anonymization remains in its infancy, leaving ample room for advancements in both fundamental understanding and applied techniques. Below are some avenues we consider promising for future exploration:

\begin{itemize}
    \item \textbf{Understanding anomalies:} Investigate the unexpected performance of linear SVC on the \textit{income} target at high $k$ values, potentially through comparative studies using synthetic datasets.
    \item \textbf{Optimizing algorithms:} Develop faster algorithmic implementations to accommodate larger datasets in real time, thereby improving generalization and allowing users to make more informed interactive decisions.
    \item \textbf{Expert-driven experiments:} Incorporate domain-specific knowledge, such as medical research, and collaborate with professionals to test whether expert input enhances both human bias and iML performance.
    \item \textbf{Engaging designs:} Explore gamified or socially-driven applications to motivate non-expert users, potentially yielding improved outcomes even in less specialized tasks.
    \item \textbf{Expanding data formats:} Investigate the application of iML to unstructured and semi-structured data, such as audio, video, and \textit{omics} datasets. Visual data, in particular, offers significant potential for medical applications due to human efficiency in image processing.
\end{itemize}

\section{Conclusion}
\label{sect:conclusion}

In this study, we explored the growing significance of privacy-conscious data management and proposed a novel interactive framework that leverages human expertise for anonymization tasks through Interactive Machine Learning (iML). Our framework combines the computational efficiency of algorithmic anonymization with the nuanced decision-making abilities of human intuition, offering a dynamic approach to preserving data utility while safeguarding privacy.

Through an experiment involving clustering data points according to user-defined preferences for attribute retention, we demonstrated the practical implications of iML in classifying anonymized personal data into critical categories such as \textit{marital status}, \textit{education level}, and \textit{income}. Our results highlighted the following key takeaways:
\begin{itemize}
    \item \textbf{Human-AI Synergy:} Human input during the anonymization process significantly improved the balance between privacy and utility, especially in scenarios where domain expertise influenced the weighting of quasi-identifiers. This suggests that human expertise can serve as a valuable complement to algorithmic processes.
    \item \textbf{Performance Stability Across Scenarios:} Despite varying levels of data generalization, iML maintained classifier performance closer to that of non-anonymized datasets, outperforming baseline methods such as equal and biased weight configurations.
    \item \textbf{Scalability of iML:} While our experiment focused on moderate-sized datasets, the framework’s modular nature indicates scalability to larger datasets with appropriate computational resources. Further, the browser-based implementation ensures accessibility across diverse user groups.
\end{itemize}

Technical Considerations
Our findings also revealed critical technical considerations for future iML-based anonymization studies:
\begin{enumerate}
    \item \textbf{Attribute Interdependencies:} Attributes often exhibit complex interdependencies (e.g., \textit{income} correlates with \textit{education level}), which can influence both utility and privacy metrics. Incorporating these interdependencies into weight configurations could further optimize outcomes.
    \item \textbf{Dimensionality Reduction:} For high-dimensional datasets, techniques such as Principal Component Analysis (PCA) or autoencoders could simplify user interactions while preserving meaningful patterns in the data.
    \item \textbf{Usability of Interfaces:} The design of the iML interface plays a pivotal role in user engagement. Our experiment demonstrated that intuitive clustering and real-time visual feedback encouraged user participation, but more sophisticated interfaces (e.g., augmented with heatmaps or confidence scores) could further enhance effectiveness.
    \item \textbf{Algorithmic Enhancements:} While \textit{SaNGreeA} proved effective in our study, integrating modern clustering techniques, such as density-based methods or neural clustering frameworks, might improve the scalability and adaptability of the system.
\end{enumerate}

Future Directions
The study opens several promising avenues for future research:
\begin{itemize}
    \item \textbf{Dynamic Weight Adjustment:} Incorporating adaptive mechanisms where the system learns from user input over time to refine attribute importance dynamically.
    \item \textbf{Real-Time Feedback on Privacy Risks:} Integrating metrics such as Differential Privacy or k-anonymity violations in real-time to help users understand the trade-offs during the anonymization process.
    \item \textbf{Domain-Specific Applications:} Testing the framework on diverse datasets, such as healthcare records, financial data, or social media logs, to evaluate its generalizability and effectiveness in domain-specific scenarios.
    \item \textbf{Multi-Objective Optimization:} Extending the framework to accommodate multi-objective optimization for balancing privacy, utility, and fairness in the anonymization process.
\end{itemize}

In conclusion, this study underscores the transformative potential of interactive approaches in data anonymization. By bridging the gap between algorithmic precision and human intuition, iML emerges as a promising paradigm for privacy-preserving data management. While our results validate the effectiveness of this framework, they also highlight the complexities involved, suggesting that advancing iML will require interdisciplinary collaboration across machine learning, human-computer interaction, and data ethics. As privacy concerns become increasingly critical in the era of big data, frameworks like ours represent a step toward responsible and transparent data practices.

\bibliographystyle{IEEEtran}
\bibliography{references}

\end{document}